\documentstyle[12pt,epsf]{article}

\begin{document}
\thispagestyle{empty}

\font\fortssbx=cmssbx10 scaled \magstep2
\hbox to \hsize{
{\fortssbx University of Wisconsin - Madison}
\hfill$\vcenter{\hbox{\bf MADPH-96-981}
                \hbox{December 1996}}$ }

\vspace{.25in}

\begin{center}
{\large\bf Status of Neutrino Astronomy:\\ The Quest for Kilometer-Scale Instruments}\\[3mm]
Francis Halzen
\end{center}

{\small\narrower\noindent
This is a (very) personal attempt to summarize the status of neutrino astronomy: its scientific motivations, our understanding of natural water and ice as particle detectors and, finally, the detector technology.
\par}

\section{Scientific Motivation}

The observed photon spectrum of conventional astronomy extends to energies in excess of 10~TeV, perhaps as high as 50~TeV. Astronomy can be done, at much higher energy, by measuring the arrival directions of cosmic rays with energy in excess of 100~EeV where they point back at their sources. At such energies their gyroradii exceed the size of our galaxy. More than 6 orders of magnitude in photon energy, or wavelength, are left unexplored. Neutrino telescopes have been conceived to fill this gap. Given the history of astronomy, it is difficult to imagine that this will be done without making major, and most likely totally surprising, discoveries. The 18 orders of magnitude in wavelength, from radio-waves to GeV gamma rays, are indeed sprinkled with unexpected discoveries. Neutrinos have the further advantage that, unlike photons of TeV energy and beyond, they are not absorbed by interstellar light. They can, in principle, reach us from the edge of the Universe. If, for instance, the sources of the high energy cosmic rays are well beyond the Virgo cluster, the photon window for their exploration closes above 100~TeV.

Speculations that the highest energy photons and protons are produced by cosmic accelerators powered by the supermassive black holes at the center of active galaxies can be used to estimate the effective volume of a neutrino telescope. The answer is that a 1~km$^3$ detector is required to detect the neutrinos accompanying the highest energy cosmic rays\cite{halzen}. This estimate finds further support when studying the other diverse scientific missions of such an instrument which touch astronomy, astrophysics, cosmology, cosmic ray and particle physics. Kilometer-scale detectors are required, according to the best available estimates, to have a good chance of finding the accelerators of the highest energy cosmic rays, of improving significantly the searches for cold dark matter particles or WIMPs, and to search meaningfully for the cosmic sources of gamma ray bursts.  In order to explore the plausible region of astronomical parameter space for these fascinating cosmic enigmas, a high energy neutrino telescope must contain several thousand optical detection modules in a volume of order 1~kilometer on a side. Model building suggests that some, and most likely the most exciting, discoveries may be within reach of much smaller detectors with effective telescope area of order $10^4\rm\,m^2$\cite{halzen}; see Fig.~1. Experience with small detectors is, in any case, an important intermediate step and early indications are that a kilometer-scale detector could be constructed in 5 years using existing technologies.

\begin{figure}[t]
\centering
\hspace{0in}\epsfxsize=5in\epsffile{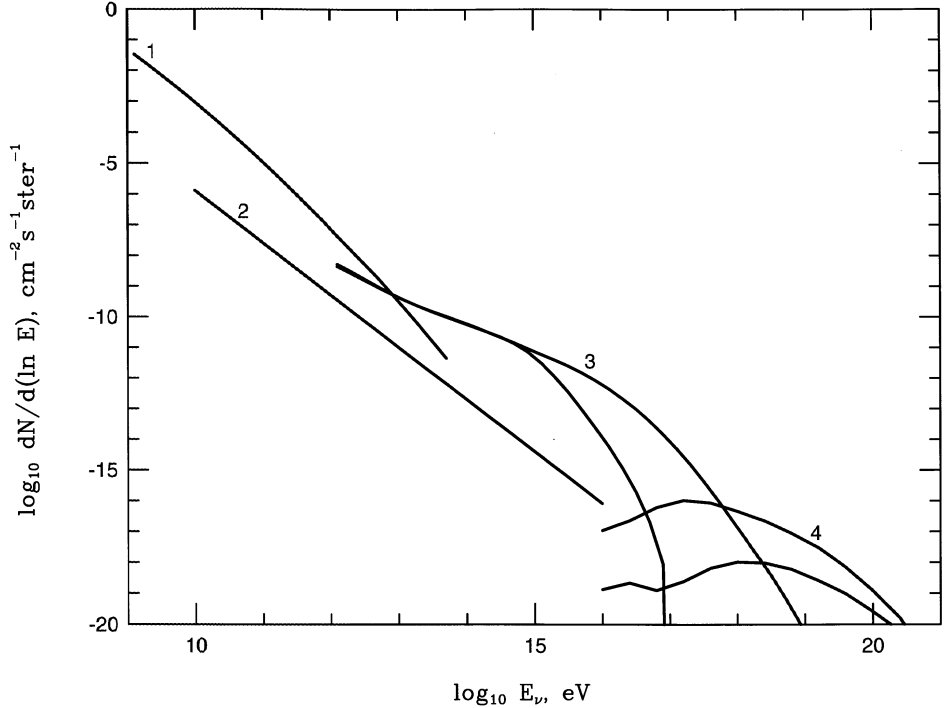}

\caption{Summary of isotropic neutrino fluxes
of energy above 1 GeV. (1)~atmospheric neutrinos; (2)~diffuse galactic neutrinos; (3)~diffuse extragalactic
neutrinos --- maximum and minimum predictions; (4)~cosmological neutrinos --- maximum and minimum predictions. From Ref.~\protect\cite{halzen}}
\end{figure}

Consisting of several thousand optical modules (OM: pressure vessel containing a conventional photomultiplier tube and, possibly, data acquisition electronics) deployed in natural water or ice, even the ultimate scope of these detectors is similar to that of the Superkamiokande or SNO solar neutrino experiments\cite{sn}. Being optimized for large effective area rather than low threshold (GeV or more, rather than MeV), they are complementary to these detectors. The challenge to deploy the components in an unfriendly environment is, however, considerable. With a price tag which may be as low as a relatively cheap fixed-target experiment at an accelerator, but could be as high as that of a LHC detector, this must be one of the best motivated large-scale scientific endeavors ever.

As for conventional telescopes, at least two are required to cover the sky. As with particle physics collider experiments, it is very advantageous to explore a new frontier with two or more instruments, preferably using different techniques. This goal may be achieved by exploiting the parallel efforts to use natural water and ice as a Cherenkov medium for particle detection.

\section{Detection Techniques\rm\protect\cite{milla}}

One can picture a neutrino telescope as a collection of strings (actually cables transmitting the signals) spaced by a distance $d_{\rm string}$ of several tens of meters. The OMs are deployed as beads on the strings and are separated by a vertical distance $d_{\rm OM}$ of order ten meters. The arrival times and number of Cherenkov photons reaching each OM are recorded in order to:

\renewcommand{\theenumi}{{\it\roman{enumi}}}
\begin{enumerate}

\item Map the Cherenkov cone radiated by muons produced by muon-neutrinos interacting inside or near the detector. At TeV energy and above the direction of the muon and incident neutrino are aligned to better than 1$^\circ$, making astronomy possible. 

\item Map the Cherenkov light made by (mostly) electromagnetic showers initiated by electron-neutrinos. One may possibly identify PeV $\tau$-neutrinos by separating the showers associated with the production and decay of the secondary $\tau$-lepton. Cosmic electron neutrinos may also be detected by the production of real weak intermediate bosons in interactions with atomic electrons: $\bar\nu_e e^-\to W^-$.

\item Detect cosmic ray muons as well as muons which originate in atmospheric cascades initiated by $\gamma$-rays from point sources.

\item Detect bursts of MeV neutrinos. The very low background counting rate of OMs deployed in the extremely transparent, sterile ice is increased by a statistically significant amount by additional signals produced by bursts of MeV neutrinos, e.g.\ from a supernova or a gamma ray burst. A stellar collapse at the center of our galaxy will be observed with good statistical significance and the time profile of the neutrino emission determined with good statistics. Relative timing of neutrinos and gamma rays from a cosmological gamma ray burst can determine the neutrino mass with a precision covering the range implied by the solar anomaly. We will not discuss the method any further. The detector capability is determined only by the absorption length of the Cherenkov medium and the background counting rate of the OMs. As we will see further on, ice has a double advantage here.

\end{enumerate}

There has been a heightened level of activity in this field in the last 3 years. In the Spring of 1993, using the frozen ice as a platform for easy deployment, the Lake Baikal group deployed a small telescope consisting of 36 optical modules. They plan to complete the detector consisting of 200 optical modules by 1998. The Russian-German collaboration has recently recorded the first candidate neutrino events.

Deep ocean water should be superior in optical quality to that in Lake Baikal. Two collaborations, ANTARES (France) and NESTOR (Greece), are developing the infrastructure and technologies for the deployment of neutrino telescopes in the Mediterranean basin. It is only fitting to also recall the pioneering role played by the now defunct DUMAND\cite{learned} experiment in Hawaii. They made key conceptual and technological contributions to this field.

The AMANDA project's goal is to commission a neutrino telescope by using natural Antarctic ice as a particle detector. With hot water, holes are melted into the 3~kilometer thick ice sheet at the South Pole into which the OMs are frozen. In Austral summer 93-94 the AMANDA A detector of 80 OMs was deployed on 4 strings positioned between 810~m and 1000~m; see Fig.~2. A deep array, AMANDA B, has been partially put in place below 1500~m. The first 4 deep strings of 20 OMs, separated by 20~m along the string, have been operating since the 95-96 season of Antarctic summer. AMANDA B will be completed with 7 strings of 36 OMs each in the 96-97 season. Within a few months close to 400 OMs will provide the collaboration with a first opportunity to do science. In the following season  strings of kilometer length will be deployed with the goal to commission a kilometer-scale detector over a period of approximately 5 years.

\begin{figure}[h]
\centering
\hspace{0in}\epsfxsize=5.2in\epsffile{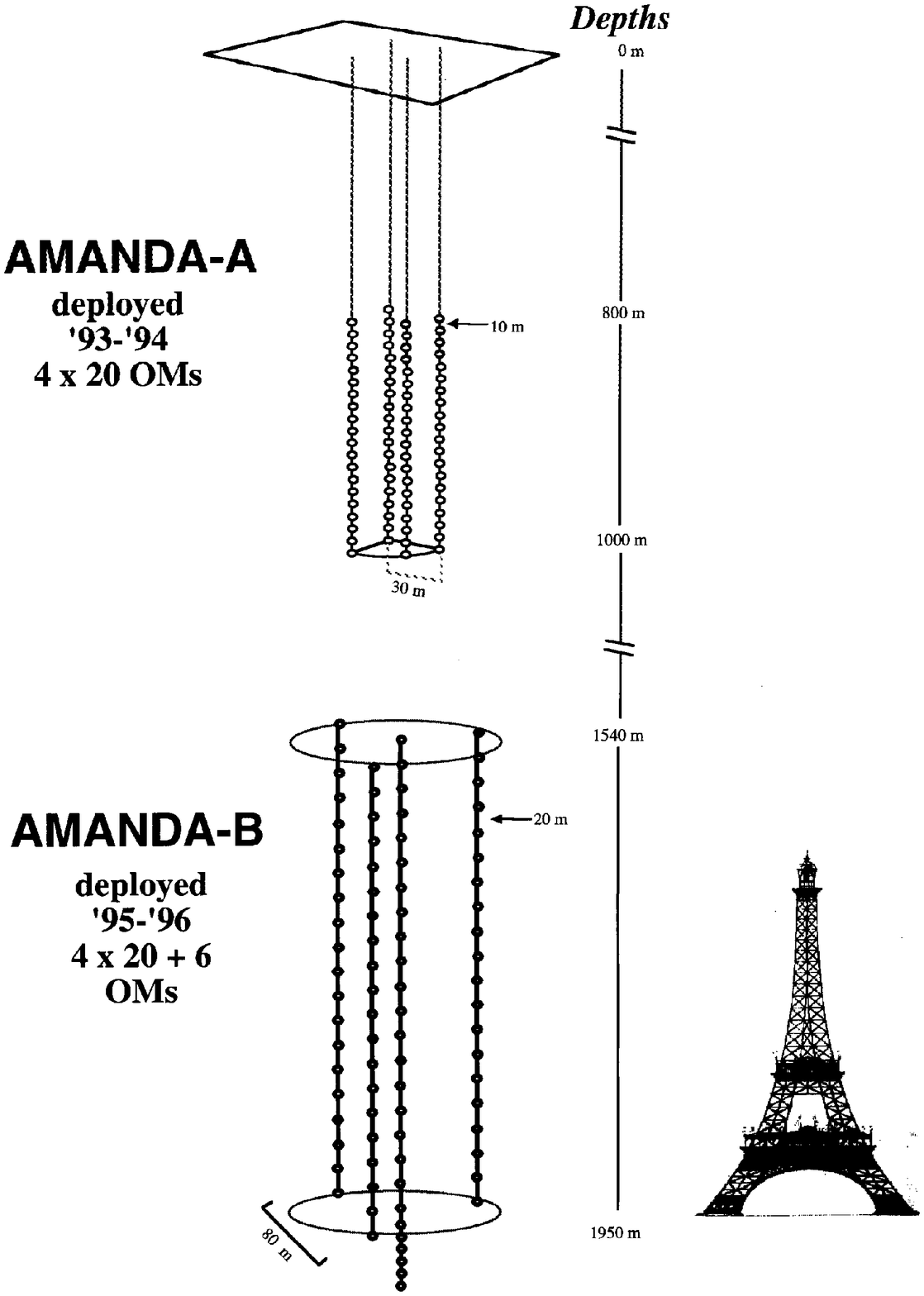}

\caption{}
\end{figure}

In order to operate these instruments one must understand {\em how} light propagates in the Cherenkov medium. Note, however, that it is not essential to understand {\em why} light propagates in the observed manner, the details of which may be very complicated. How light propagates is determined by independent methods, e.g.\ by studying light emitted from sources frozen in with each OM which can be pulsed by a laser. The AMANDA experiment is also calibrated on a beam of muons, selected in energy and direction, by the SPASE air shower array.

The simplest parametrization of the propagation of light in a medium is made in terms of an absorption length $\lambda_{\rm abs}$ and an effective scattering length $\lambda_{\rm sc}$ (geometric scattering length divided by ($1-\tau$), where $\tau$ is the average scattering angle in each interaction. A more detailed representation of the actual angular distribution of scattered light will become necessary as more refined experimental information accumulates. A term frequently used in ocean optics is attenuation length, the harmonic sum of absorption length and scattering length.

\begin{itemize}

\item {\bf Water}. The optical properties of water vary with the color of the light. The quantum efficiency of photomultiplier tubes peaks near 400~nm. The relevant colors are somewhat below this value because of the falling spectrum of blue Cherenkov photons. At the Dumand site the attenuation length at 400~nm has been measured to be $16.7\pm1.5$~m. It peaks in the vicinity of 500~nm at about 40$\sim$50~m. Similar results have been obtained at the Nestor site. The attenuation length is roughly a factor of 2 smaller in Lake Baikal, where the scattering length, in excess of 100~m, has been measured separately. There are indications that its value is similar, likely larger, in deep ocean water. Long-time measurements in Lake Baikal exhibit seasonal dependence of the optical properties of the lake\cite{jpl}.

\item {\bf Ice}.  At 850~m depth, in the ultrapure ice formed just after the most recent ice age, a variety of in-situ measurements of the optical properties of South Pole ice revealed an absorption length exceeding 200~m for wavelengths below 425~nm, more than one order of magnitude larger than laboratory ice\cite{science}. Unfortunately, remnant bubbles, much larger in size than expected, limit the  scattering length to tens of centimeters, preventing the reconstruction of muon tracks. (With this set-back nature did however provide the collaboration with the hint that scattering can be exploited to measure energy; more about this later). Measurements for the AMANDA B strings at 1500 to 1900 m have confirmed the large absorption lengths observed at 0.8 to 1 kilometer (AMANDA A) and indicate a  scattering length which is 2 orders of magnitude larger than for AMANDA A. The optical properties of the ice  do not vary over the interval 1950 to 1850 m for which data now exist. Light from a 337~nm laser reveals a scattering length of 25~m at that wavelength; the recorded muon tracks do, however, indicate that at least some of the Cherenkov photons propagate much greater distances. More about this later as well. Data for AMANDA B taken between 337 and about 500 nm show that scattering is independent of wavelength and absorption depends only weakly on wavelength.

\end{itemize}

In summary, indications are that water and ice have complementary optical properties: while the attenuation lengths are comparable, scattering limits the performance in ice whereas absorption limits the performance in water. From the point of view of optics only, a neutrino telescope can be built in either medium with $d_{\rm OM}\simeq 10{\sim}20$~m, $d_{\rm string} \simeq\lambda_{\rm attenuation} \simeq 30{\sim}60$~m. The fine-tuning of these distances depends on one's priority of science goals, especially because the spacings (along with the trigger conditions) control the threshold of the detector.

Besides optics, there are other important factors determining the performance of a detector: the cosmic ray muon background and the background counting rate in the OMs. The capability to shield cosmic ray background is limited by a maximal depth of 3~km of the ice sheets, while a detector can be completely immersed below 3~km of water at the best Mediterranean sites, e.g.\ the location of the NESTOR experiment. All experiments use the earth as a filter: muons coming up through the earth are of neutrino origin, down-going muons are cosmic ray background. The OM background counting rate from natural radioactivity has been measured to be 60~kHz in all aquatic sites, while it is only a few 100~Hz, on average, for the AMANDA modules presently operating in 1.5--2~km deep, sterile South Pole ice.

The most important factor may, in the end, be cost. The technology must be cheap in order to scale the detector up to kilometer dimensions. There is, in principle, no reason why a neutrino telescope cannot be built using existing ``accelerator-based'' technology. A simple estimate anticipates a price tag of the order of 10~billion US dollars. We know from experience that this is not a good idea. We must develop techniques which are more cost-effective by 2 orders of magnitude. The cost is not only associated with the hardware, but with the deployment and operation of the detector in a hostile environment such as the deep ocean, or a remote Antarctic station.

Before proceeding, it should be pointed out that the challenge of constructing a small prototype is different, and, in fact,  far greater, than for building a kilometer-scale instrument. We will discuss prototype and kilometer-scale detectors sequentially.

\section{First-Generation Neutrino Telescopes}

First-generation detectors are designed to reach an effective telescope area of several times $10^4$~m~$^2$, maybe less for the Baikal experiment. Because the instrumented volume is limited, these detectors stretch ``effective volume" by attempting to reconstruct the Cherenkov cone of muon tracks far outside the detector. With their relatively small spacings the Baikal and NESTOR detectors have a threshold of a few GeV, i.e.\ a typical event consists of a short muon track of about 10~m length, typically outside the detector. Recall that a muon travels approximately 5~m per GeV of energy. The Baikal group has actually met the challenge of catching such events in a background of down-going cosmic ray muons, a million times more numerous. They reconstruct neutrinos to degree accuracy with as few as 4 OMs reporting. Their effective volume is however limited --- tracks far outside the detector are beyond reach. A deeper ocean detector should benefit from a reduction in background and better optical quality of the water.

The remote deployment of OMs in water requires local data acquisition systems, housed inside or near the OMs, which transmit the information over fiber optic cable to shore. Last April the Baikal collaboration deployed 96 OMs. After several months, all but two were stably operating, a major improvement over previous campaigns. Subsequently a string had to be turned off because of data transmission problems. It is hopefully not too early to conclude that the underwater technology is (almost) under control. Compared to ocean experiments the Baikal experiment is, however, in some respects special: they can deploy from the stable ice platform that covers the lake several months of the year, there are no waves, they can service the experiment yearly and, they deploy OMs only to 1~km depth.

The AMANDA philosophy is different\cite{heidelberg}. A much larger volume of ice is instrumented with a large number of smaller OMs, equipped with 8 rather than 20~inch photomultipliers. Spacings are, on average, much larger; see Fig.~2. Only tracks of 100~m length or more are reconstructed resulting in an increased threshold of 20~GeV. In the end larger volume and higher threshold compensate and the number of atmospheric neutrinos detected is similar. It is also worth mentioning that the increased background at 1.5--2~km depth is partially overcome by pointing all OMs down. 

Because of the long absorption length a typical atmospheric neutrino event will trigger $\sim$~20 optical modules, yielding ample information to reject the background and reconstruct the track. Degree accuracy is achieved by a likelihood track fit which requires that more than 6$\sim$8 photons are not scattered in events with more than 15$\sim$20 OMs reporting. Because of the excellent transparency, such cuts can be made without the dramatic loss in effective volume they would engender in a water detector. Even before reconstruction of muon tracks is attempted, AMANDA events, acquired at a steady 100~Hz rate, can be reduced to roughly 10$^2$ events per day using simple, on-line, timing cuts which require that time flows up, rather than down in an event record (particle physicists would call this a level-3 trigger). This set of events consists of showers, showering muons and candidate neutrino events; a neutrino event is shown in Fig.~3. Notice the large number of photons produced in the very transparent deep ice; the event is not atypical. 

\begin{figure}
\centering
\hspace{0in}\epsfxsize=5in\epsffile{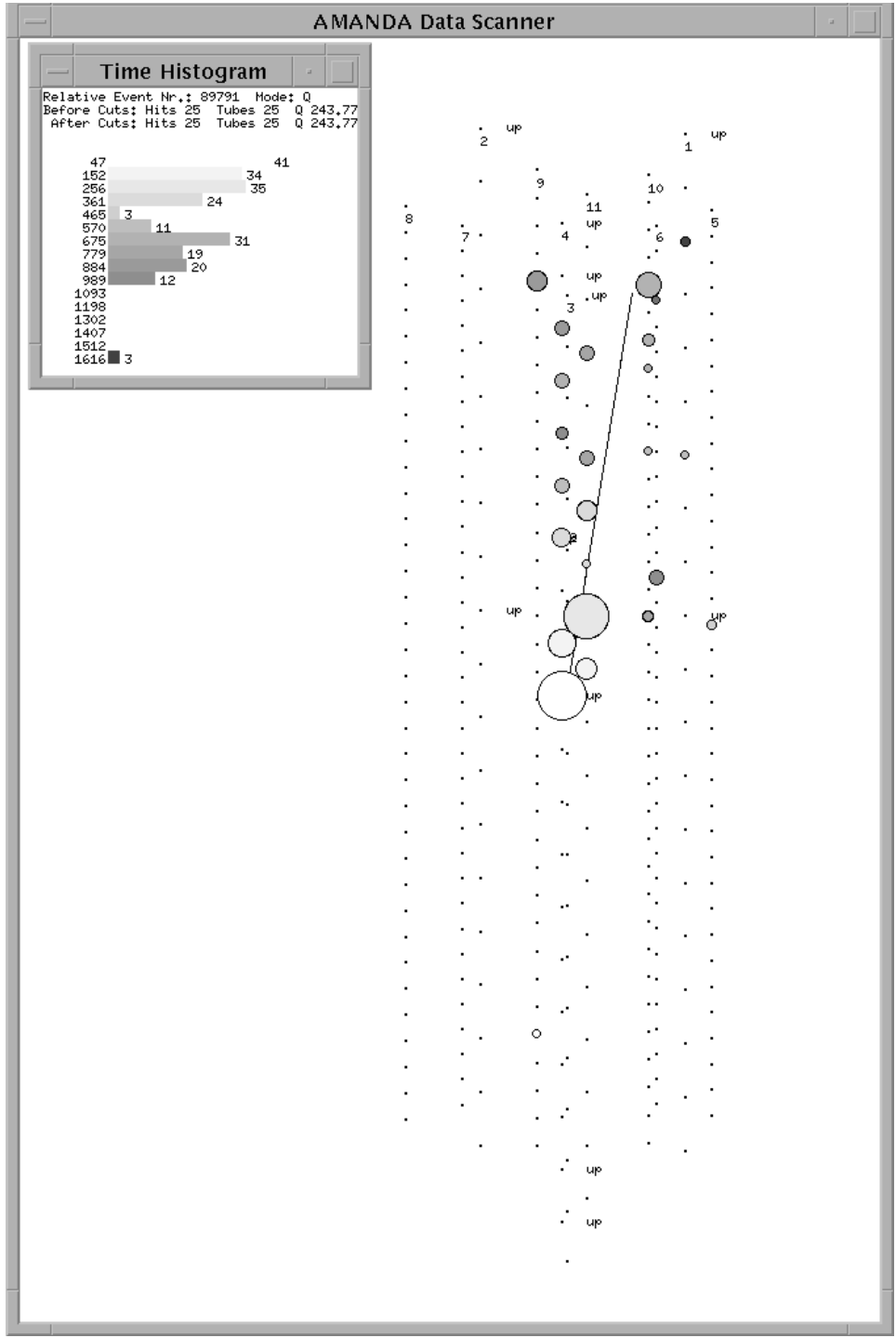}

\caption{Monte Carlo simulation of a typical AMANDA B event. Dots indicate triggered OMs. Their size reflects the number of photons, the shading indicates time.}
\end{figure}

With hot water, holes are melted into the 3~kilometer thick ice sheet at the South Pole. Optical modules, consisting of a photomultiplier in a pressure vessel and nothing else, are frozen into the ice. Hot water drilling progresses 1~centimeter every second; 2-kilometer deep holes are drilled in 4~days. Less than 10\% of the OMs fail during deployment, none has failed subsequently. The possibility to operate the detector with a data acquisition system placed right on top of the detector allows for a simple and non-hierarchical system where each OM is connected to the surface by its own cable. This provides high reliability against single point failure and makes evolutionary upgrades of the instrument possible. Furthermore, ice is a totally sterile medium devoid of radioactivity. Background counting rates in the OMs are extremely low: only a few hundred Hz for the deep AMANDA detector. The experiment can therefore be triggered by off-the-shelf electronics. It was indeed the low cost which represented the initial motivation for the experiment. The proponents argued that ice represented a competitive alternative to water even though the absorption length of blue light, relevant to the propagation of Cherenkov photons, was thought to be only 8~meters based on laboratory experiments. Nature has been more kind.

 As part of the evolving detector technology OM signals will in the future be transmitted over twisted pair cable which in tests yields a factor 3 improvement in both rise-time and amplitude compared to coax. The smaller diameter of the cable makes possible the increased number of OMs per string. A new technology will also be tested where the dynode signal of the photomultiplier drives a laser or a fast optical LED which transmits the analog signal over fiber optic cable. It is likely that this method, if proven reliable, will be the AMANDA technology of the future. A pair of digital modules, produced at the Jet Propulsion Laboratory of NASA, will also be tested as part of the next deployment.

In summary, the techniques are on paper (actually by Monte Carlo) competitive, although the AMANDA technology is simpler, more cost-effective and, it has been demonstrated. This is exploratory science however and the issues are sufficiently complex that it would be prudent to weigh the relative merits of the techniques on the basis of data from first-generation detectors, not Monte Carlo.

\section{Speculations on a Kilometer-Scale Detector}

The present experiments can be used as components of a modular detector reaching a few times 0.1~km$^3$ volume, possibly more. Alternatively, one may envision detectors where OMs are uniformly distributed in a teraliter (km$^3$) of water or ice. Conceptually there is no real challenge. Muon tracks are now hundreds of meters in length and can be reconstructed with degree accuracy. Up-down discrimination of tracks far outside the detector is no longer necessary --- one can limit the detector volume essentially to the instrumented volume.

The threshold (and cost) of such an instrument is determined by the number of OMs and is of order 1~PeV for a thousand 20~inch OMs\cite{stenger} deployed in ocean water below 4~kilometers. Stenger's simulations illustrate new challenges of kilometer-size detectors. Even after removing all the 1~p.e.\ signals which are dominated by background, half of the ${\sim}10^2$ OM signals from a PeV muon are produced by potassium decay in natural water. Triggering now becomes the real challenge and the up/down discrimination problem may rear its ugly head in the confusion created by the large number of noise hits. Stenger shows that the latter problem can be handled. Still, my guess is that the major issue for water detectors will be cost. Only when the multiple R\&D efforts, presently in progress, deliver reliable deployment schemes for constructing detectors with a lifetime of several decades, can this question be answered.

What about ice? This is pioneering science --- surprises are routine. The two most important discoveries after operating AMANDA A and B for 2 years and 10 months, respectively, are that neutrino energy can be measured and that bursts of MeV-neutrinos, although well below the nominal threshold of the neutrino telescope, can be detected. The capability to operate a detector with linear energy response is relevant to the consideration of large detectors: a kilometer-scale ice detector is expected to operate as a directional Cherenkov detector as well as a total absorption calorimeter. The energy measurement is critical because, once it can be established that neutrino energy exceeds 10~TeV, cosmic origin of the signal is established because the atmospheric background rate is negligible as a result of the steeply falling spectrum; see Fig.~1. The problem of reconstructing muons in a kilometer-scale detector has been assessed experimentally by i) studying muon tracks registered in both the 1 and 2 kilometer-deep detectors and ii) investigating linear energy response in AMANDA~A. We discuss this next.

\renewcommand{\thefigure}{\arabic{figure}a}
\begin{figure}
\centering
\hspace{0in}\epsfysize=6.5in\epsffile{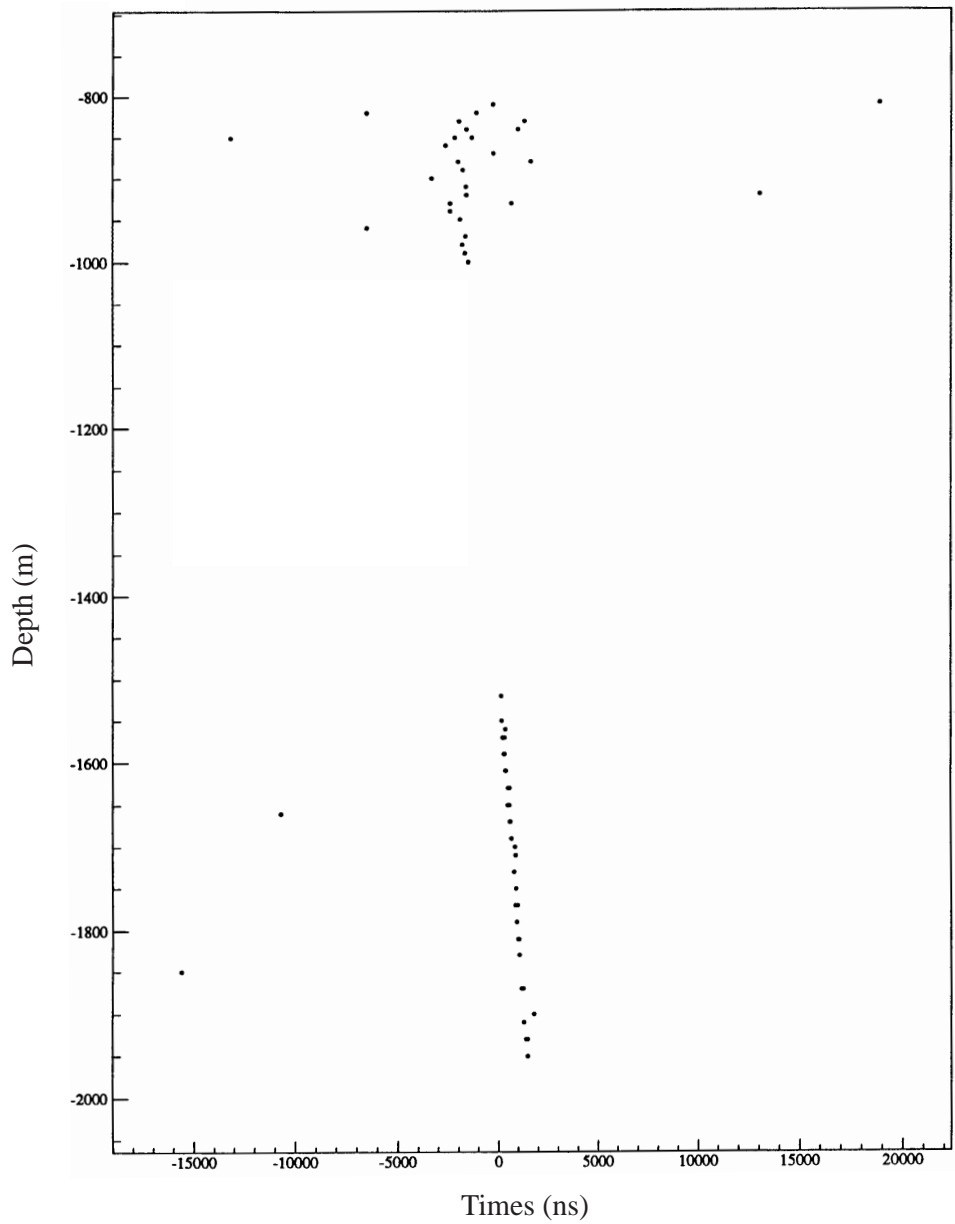}

\caption{Cosmic ray muon track triggered by both AMANDA A and B. Trigger times of the optical modules are shown as a function of depth. The diagram shows the diffusion of the track by bubbles above 1~km depth. Early and late hits, not associated with the track, are photomultiplier noise.}
\end{figure}

\addtocounter{figure}{-1}\renewcommand{\thefigure}{\arabic{figure}b}
\begin{figure}
\centering
\hspace{0in}\epsfysize=6.5in\epsffile{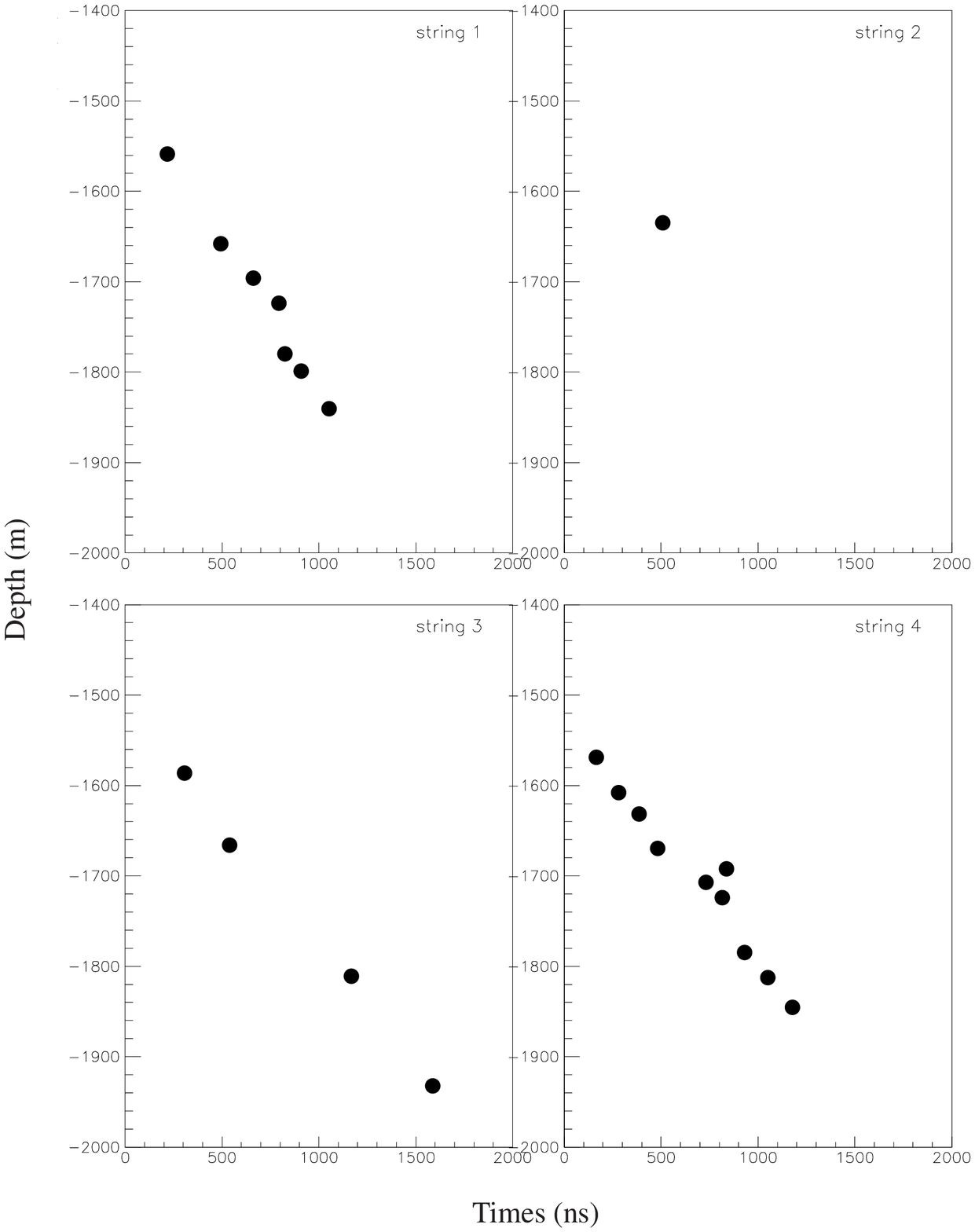}

\caption{Cosmic ray muon track triggered by both AMANDA A and B. Trigger times are shown separately for each string in the deep detector. In this event the muon mostly triggers OMs on strings 1 and 4 which are separated by 79.5~m. }
\end{figure}

Coincidences between AMANDA A and B are triggered at a rate of 0.1~Hz. Every 10 seconds a muon is tracked over 1.2 kilometer; a typical event is shown in Fig.~4. Below 1500~m the vertical muon triggers 2 strings separated by 79.5~m. The distance along the Cherenkov cone is over 100~m, yet, despite some evidence of scattering, the speed of light propagation of the track can be readily identified. We have already analysed $5 \times 10^5$  A-B coincidences with no evidence for any source of misreconstructed background events. Reconstruction of such tracks to a degree in bubble-free ice should not represent a challenge.

The shallow part of the AMANDA detector has been operated for over 2 years as a shower detector and the first few months of data analysed. With a scattering length of order 50~cm, AMANDA A is an adequate total absorption calorimeter despite its limited size. Bubbles diffuse and contain the shower light very effectively. The shower can be mapped and its energy measured provided the point of origin is within 50~m of the instrumented ice. Candidate atmospheric neutrino events have been identified in the range of 100~GeV to 1~PeV. A measurement of the spectrum requires a detailed understanding of all systematics. This is a considerable challenge, especially because of the limited amplitude information in the present detector. In bubbly ice the directional information is essentially lost. AMANDA~A is the ideal prototype to develop the methods for measuring energy. It is, with its reduced detector dimension and scattering length, a scale model of a kilometer-cubed detector.

Ice is an ideal Cherenkov medium for a kilometer-scale neutrino detector. Simulations show that a PeV shower, initiated by a $\nu_e$ or $\nu_\tau$ or by catastrophic energy loss of a muon, travels over 500~meters in the very transparent ice. At this point the shower is isotropized by scattering and its energy can be determined. The first photon which arrives at an OM is not scattered by the ice provided it is within 200~meters from the point of origin of the shower. This will supply directional information. As already pointed out, once one establishes that the energy of the event exceeds those of the highest energy cosmic ray muons registered by the detector, the up-down discrimination problem is automatically solved. 

The AMANDA collaboration has argued that they can deploy a high threshold, perhaps 10~TeV, kilometer-scale detector for 30~million dollars, including all the cost of South Pole logistics. It consists of 80 strings with 50 OM on each, for a total of 4000 small 8~inch (perhaps 12~inch) photomultipliers. The relatively low cost is not too surprising if one recalls that cost represented the major initial motivation for the experiment. The proponents argued that ice represented a competitive alternative to water even though the absorption length of blue light was thought to be only 8~meters based on laboratory experiments.

\section{Conclusions}

The hope that complementary water and ice detectors will soon probe Northern and Southern hemispheres for cosmic sources of neutrinos is not unrealistic. Depending on the results, high threshold detectors, constructed in the next decade, can subsequently be expanded or, alternatively, back-filled with additional OMs to reach lower thresholds. 

\section*{Acknowledgments}
This research was supported in part by the U.S.~Department of Energy under Grant No.~DE-FG02-95ER40896 and in part by the University of Wisconsin Research Committee with funds granted by the Wisconsin Alumni Research Foundation.

\end{document}